\documentclass[pre,aps,10pt,floatfix,twocolumn,showkeys,superscriptaddress]{revtex4-2} 
\usepackage{color, soul, float, graphicx} 
\usepackage{amsmath, amsthm, amssymb, mathtools, siunitx}
\usepackage{multirow} 
\usepackage[normalem]{ulem}
\begin{document} 
\title{Conformational landscape of long semiflexible linear and ring polymers near attractive surfaces} 
\author{Kamal Tripathi} 
\email{kamalt@imsc.res.in}
\affiliation{The Institute of Mathematical Sciences, C.I.T. Campus, Taramani, Chennai 600113, India}
\author{Satyavani Vemparala} 
\affiliation{The Institute of Mathematical Sciences, C.I.T. Campus, Taramani, Chennai 600113, India}
\affiliation{Homi Bhabha National Institute, Training School Complex, Anushakti Nagar, Mumbai 400094, India}

\begin{abstract}
Conformations of a crowded neutral semiflexible polymer under confinement near an attractive wall are studied via coarse-grained simulations. We study the effects of the interplay of the length of the polymer, bending rigidity, and the repulsive crowder density on such equilibrium semiflexible polymer conformations. The length of the polymer dictates the number of distinct conformations of the adsorbed semiflexible polymer, suggesting that previous studies of short polymers are limited. The crowder density is shown to effectively reduce the bending rigidity of the polymers and at the highest crowder density considered, a crumpled wall-adsorbed polymer conformation is seen regardless of the semiflexibility. In addition, we study the role of the topology of the semiflexible polymer by comparing the results of linear semiflexible polymers with those of ring polymers. The conformational landscape of crowded ring polymers shows less diversity than that of a linear polymer and crowders are seen to affect the ring polymers differently than their linear counterparts.
\end{abstract}
\maketitle
%%%%%%%%%%%%%%%% INTRODUCTION%%%%%%%%%%%%%%%%%%%%%%%%%%%%%%%%%%
\section{Introduction} 
Many semiflexible biopolymers such as DNA, actin, and microtubules are some of the main components inside a biological cell and are often seen in a condensed state, which could be in either compaction or bundled up polymers~\cite{alberts2008molecular, rafelski2004crawling, schnauss2016semiflexible}. Biopolymers have a wide range of persistence lengths ranging from a few nanometers (DNA) to a few millimeters (microtubules)~\cite{alberts2008molecular, phillips2012physical}. These biopolymers are most often confined and are in a crowded environment and this necessitates an understanding of the role of crowders and confinement on the conformational landscape of such semiflexible polymers. Further, many of these semiflexible polymers, especially in a biological context, interact with surfaces~\cite{safinya1998dna,rechendorff2009persistence,custodio2016cell,kuznetsov1998semiflexible,marenduzzo2006depletion}.

Understanding the conformational landscape of adsorbed polymers on surfaces has been of interest for a long time~\cite{de1976scaling,de1987polymers,linse2010polymer,o2003irreversibility,huber1998polymer,hershkovits2007polymer,zhu2021conformational,rajesh2002adsorption,kumar2001adsorption,venkatakrishnan2018polymer,ji1988polymer,kim2015polymer}. The adsorbing surface plays a crucial role in evolving polymer structures, which are unattainable in a bulk-like environment~\cite{tripathi2019confined, de1987polymers,rajesh2002adsorption,arkin2012structural,krawczyk2005layering}. Theoretical and numerical studies of flexible polymers~\cite{rajesh2002adsorption, bachmann2005conformational, arkin2012ground, krawczyk2005layering} suggest certain predominant phases near an attractive surface, depending on the solvent conditions:  the desorbed-extended (DE), desorbed-collapsed (DC), adsorbed-extended (AE) and adsorbed-collapsed (AC) phases. It is to be noted that most of these studies do not take into account the presence of a crowded environment, confining conditions, or flexibility of the polymer under consideration.  While there are many studies on flexible polymer adsorption on surfaces, aspects of polymer semiflexibility, crowded conditions, and the role of polymer topology on the adsorbed polymer conformations are less well studied. Earlier studies of semiflexible polymers (without crowders) with surfaces show somewhat contradictory results: for weak surface interactions, the semiflexible polymers are suggested to have a weaker binding to a flat surface than the flexible ones ~\cite{stepanow2001adsorption} while other studies ~\cite{sintes2001adsorption, kramarenko1996molecular,chauhan2021adsorption} suggest the opposite. There is also a wide range of polymer conformations seen in many computational studies. It has been suggested that the semiflexible polymers adopt simple extended chain-like structures, a sequence of alternating trains of wormlike short loops and coil-like loops forming a triple-layer structure~\cite{semenov2002adsorption, chauhan2021adsorption,ivanov2009conformational, doi:10.1021/ma970827m}. One common aspect of these studies is the short length of the polymers considered. Especially given that these studies explore the conformational behavior of semiflexible polymer, it can be envisaged that the length of the polymers can give rise to a richer variety of polymer conformations due to the interplay between persistence length and polymer length. Studies~\cite{PhysRevLett.99.198102} have shown that the shape of semiflexible polymer rings exhibits two distinct regimes depending on their flexibility \textit{i.e.,} rings with higher bending rigidity are 2D ellipses, while more flexible rings show 3D, crumpled structures. In addition, the physical properties of a polymer depend upon a number of factors and molecular topology plays a crucial role. The topology of the polymer (linear vs ring) has been shown to affect polymer conformations in confined and crowded environments~\cite{shin2014mixing,swain2019confinement,jeon2017ring,chauhan2021crowding}. Ring polymers are common in biology: E.coli's DNA and circular plasmids are notable examples of it. The curvature of the adsorbing surface is also shown to affect the polymer conformation~\cite{arkin2012structural, bachmann2005conformational, moddel2014adsorption}. 

The ubiquitous presence of crowders in systems like biological cells can have profound effects on the conformational properties of largely semiflexible biopolymers present~\cite{zhou2008macromolecular,rivas2016macromolecular,owen2019effects,kang2015effects}. Semiflexible biopolymers such as actin, DNA, etc., have been shown to undergo compaction due to the presence of both adsorbing and non-adsorbing crowder particles~\cite{kang2015unexpected, zinchenko2016dna}. While the adsorbing crowder-induced collapse is attributed to bridging-induced effective attraction within polymers~\cite{brackley2020polymer, brackley2013nonspecific}, the non-adsorbing crowder particles induce compaction via well-known depletion interaction~\cite{lau2009condensation, clarke2022depletion}. Experiments have shown that the presence of crowder particles (in the form of proteins) inside a cell can modulate the stiffness of biopolymers~\cite{claessens2006actin}, suggesting an interplay between the crowder density and effective semiflexibility of the polymers, which will eventually dictate the condensed states. Experiments have shown that the presence of crowders can lead to phase separation of semiflexible biopolymers and their eventual adsorption on surfaces~\cite{biswas2012phase}. Our earlier study~\cite{tripathi2019confined} focused on understanding the role of crowder density and confining volume on the conformational phase diagram of a long, completely flexible neutral polymer chain near both attractive as well as repulsive spherical walls. Our simulations suggested for poor solvent conditions, with the attractive interaction strength of the wall, the linear neutral polymer exhibits a transition from adsorbed globule (AG) to adsorbed layered (AL) conformation.  However, with the increasing of crowder density, the adsorbed polymer is seen to exhibit a crumpling transition, reminiscent of the SAG phase predicted in earlier theoretical study~\cite{rajesh2002adsorption}. In this study, we focus on the role of polymer length, crowder density and topology on the conformational landscape of long confined semiflexble polymers near an attractive surface. The paper is organized as follows. Section~\ref{sec:method} introduces the model and methods used in the work. Sections~\ref{sec:results} discusses the results. Section~\ref{sec:conclusions} contains our conclusions and discussions.

%%%%%%%%%%%%%%%%%%%%%%%%%%%%%%%%%%%%%%%%%%%%%%%%%%%%%%%%%%%%%%%%%%%%%%%%%%%%%%%%%%%%%%%%%%%%%%%%
%%%%%%%%%%%%%%%%%%%METHODS%%%%%%%%%%%%%%%%%%%%%%%%%%%%%%%%%%%%%%%%%%%%%%%%%%%%%%%%%%%%%%%%%%%%%%
%%%%%%%%%%%%%%%%%%%%%%%%%%%%%%%%%%%%%%%%%%%%%%%%%%%%%%%%%%%%%%%%%%%%%%%%%%%%%%%%%%%%%%%%%%%%%%%%
%\newpage
\section{Methods \label{sec:method}} 
In this study, we apply a coarse-grained molecular dynamics approach to simulate a semiflexible neutral polymer inside spherical confinement using a bead-spring model.
Crowders are modeled as spherical particles and their density is varied to mimic different crowding conditions in a spherical confinement of fixed radius. The monomers and crowders have equal size and are enclosed in a confining spherical shell of radius $R_c = 15.0$. The number of crowder particles is varied to simulate the various crowder densities The volume fraction of crowders $\phi_c$ is defined as, $\phi_c = N_c v/V_c$, where $N_c$, $v$ and $V_c$ represent the number of crowder particles, the volume of a crowder particle ($4 \pi (\sigma_c/2) ^3 /3$, where $\sigma_c$ is crowder diameter) and volume of the confining sphere respectively. The value of $\phi_c$ is varied from $0.0$ to $0.4$ in the interval of $0.05$ in the simulations.  The flexibility of the polymer is varied via the bending rigidity parameter. The length of the polymer chain $N$ is varied ($50, 400$ monomers) to understand the effects of the chain length on the conformations. We have also performed simulations of ring polymers with similar parameters to understand the effects of closed topology on the conformational properties of the semiflexible polymers.

The non-bonded interaction among the monomers, the crowders and between the monomers and crowders is modelled via shifted and truncated Lennard-Jones (LJ) potential, 
\begin{equation}
U(r_{ij}) =
\begin{cases}
4 \epsilon_{ij}\left[ \left(\dfrac{\sigma_{ij}}{r_{ij}}\right)^{12} - \left(\dfrac{\sigma_{ij}}{r_{ij}}\right)^{6} \right] + \epsilon  & r <  r_c\\
0  & r \geq r_c
\end{cases},
\label{eq:LJST}
\end{equation}
where the indices $i$ and $j$ can take either $m$ or $c$, and $m, c$ refer to the monomers and crowder particles respectively, and $r_{ij}$ represents the distance between two particles at position $r_i$ and $r_j$. The parameter $\epsilon_{ij}$ represents the strength of the interaction between the particle $i$ and $j$, and $\epsilon$ is the shift of the potential at the cutoff distance $r_c$ while $\sigma_{ij}$ is the distance at which the $U(r_{ij})$ crosses zero value. The interactions among the monomers are attractive to mimic poor solvent condition for the polymer, while the interactions between the monomers and crowders are repulsive. The interaction between the confinement wall ($w$) and a particle is modeled using truncated LJ potential and the cutoff distance of the potential depends upon whether the particle is attracted to or repelled from the wall. In this study, we have chosen monomers to have attractive interaction and crowders to have repulsive interaction with the wall. All the values of LJ potential parameters used are shown in Tab.~\ref{tab-LJ}.
\begin{table}
\centering
\caption{Table of parameters $\epsilon_{ij}$, $\sigma_{ij}$ and $r_c$ for different particle pair combinations between a monomer ($m$), a crowder ($c$) and the wall ($w$). 
\label{tab-LJ}}
\begin{ruledtabular}
\begin{tabular}{c  c  c  c}
Pair & $\epsilon_{ij}$ & $\sigma_{ij}$ & $r_c$  \\
\hline
 $m$-$m$ & 1.0 & 1.0 & 2.5\\
 $c$-$c$ & 1.0 & 1.0 & $2^{1/6}$\\ 
 $m$-$c$ & 1.0 & 1.0 & $2^{1/6}$\\ 
 $m$-$w$ & 5.0 & 1.0 & $2.5$\\
 $c$-$w$ & 1.0 & 1.0 & $2^{1/6}$\\ 
\end{tabular}
\end{ruledtabular}
\end{table}

A harmonic potential is used between nearest neighbours along the polymer chain to enforce chain connectivity. It is defined through,
\begin{equation}
U_{bond}(r_{ij}) = \frac{1}{2} K_{b} (r_{ij}-r_0)^2,
\label{eq:HarmonicPot}
\end{equation}
where $K_b$ is stiffness of the bond and is taken to be $500k_B T/\sigma_{mm}^2$ in this study. The average bond length $r_0$ for the polymer has been taken to be $1.122$.

A cosine angle potential is used to simulate the semiflexibility of the polymer. The angle potential is given by the following form,
\begin{equation}
U_{angle}(\theta) = K(1-\cos\theta),
\label{eq:CosinePot}
\end{equation} 
where $K$ and $\theta$ represent the bending rigidity of the polymer and the angle between the two consecutive bonds respectively. The values of $K$ chosen are $10, 20$ and $50$ to understand the effects of varying degrees of rigidity. 

The initial configuration of the system is constructed using a custom code written in the python programming language. The particles are then kept inside a larger spherical shell ($R = 100$) and the shell is isotropically compressed to achieve the desired radius of the confining sphere($R_c = 15$). After that, the system is equilibrated for $10^6$ time steps followed by $10^7$ time steps to generate the production data which is used for the analysis. For each set of parameters, we performed $10$ simulations, each starting from different initial conditions to ensure good statistical averaging. The equations of motion are integrated using the velocity-Verlet algorithm~\cite{verlet1967computer, swope1982computer}. The step size is taken to be $\delta t = 0.01\tau$, where $\tau = \sigma \sqrt{m/\epsilon}$, $m$, $\sigma$ and $\epsilon$ are units of mass, length and energy respectively. All simulations are performed under constant volume (V) and temperature conditions $(T = 1.0)$ using a Nos\'{e}--Hoover thermostat~\cite{nose1984unified, hoover1985canonical}. LAMMPS~\cite{plimpton1995fast} software package is used for the MD implementation. Visualization of the trajectories is done using the VMD package~\cite{HUMP96}. Custom-written tcl and python scripts are used in combination with VMD for the quantitative analysis.

%%%%%%%%%%%%%%%%%%%%%%%%%%%%%%%%%%%%%%%%%%%%%%%%%%%%%%%%%%%%%%%%%%%%%%%%%%%%%%%%%%%%%%%%%%%%%%%%%%
%%%%%%%%%%%%%%%%%%% RESULTS %%%%%%%%%%%%%%%%%%%%%%%%%%%%%%%%%%%%%%%%%%%%%%%%%%%%%%%%%%%%%%%%%%%%%%
%%%%%%%%%%%%%%%%%%%%%%%%%%%%%%%%%%%%%%%%%%%%%%%%%%%%%%%%%%%%%%%%%%%%%%%%%%%%%%%%%%%%%%%%%%%%%%%%%%
\section{Results}\label{sec:results}
\subsection{Polymer conformations in the absence of the crowders}\label{sec:A}
\begin{figure}[h]
\center
\includegraphics[width=\linewidth]{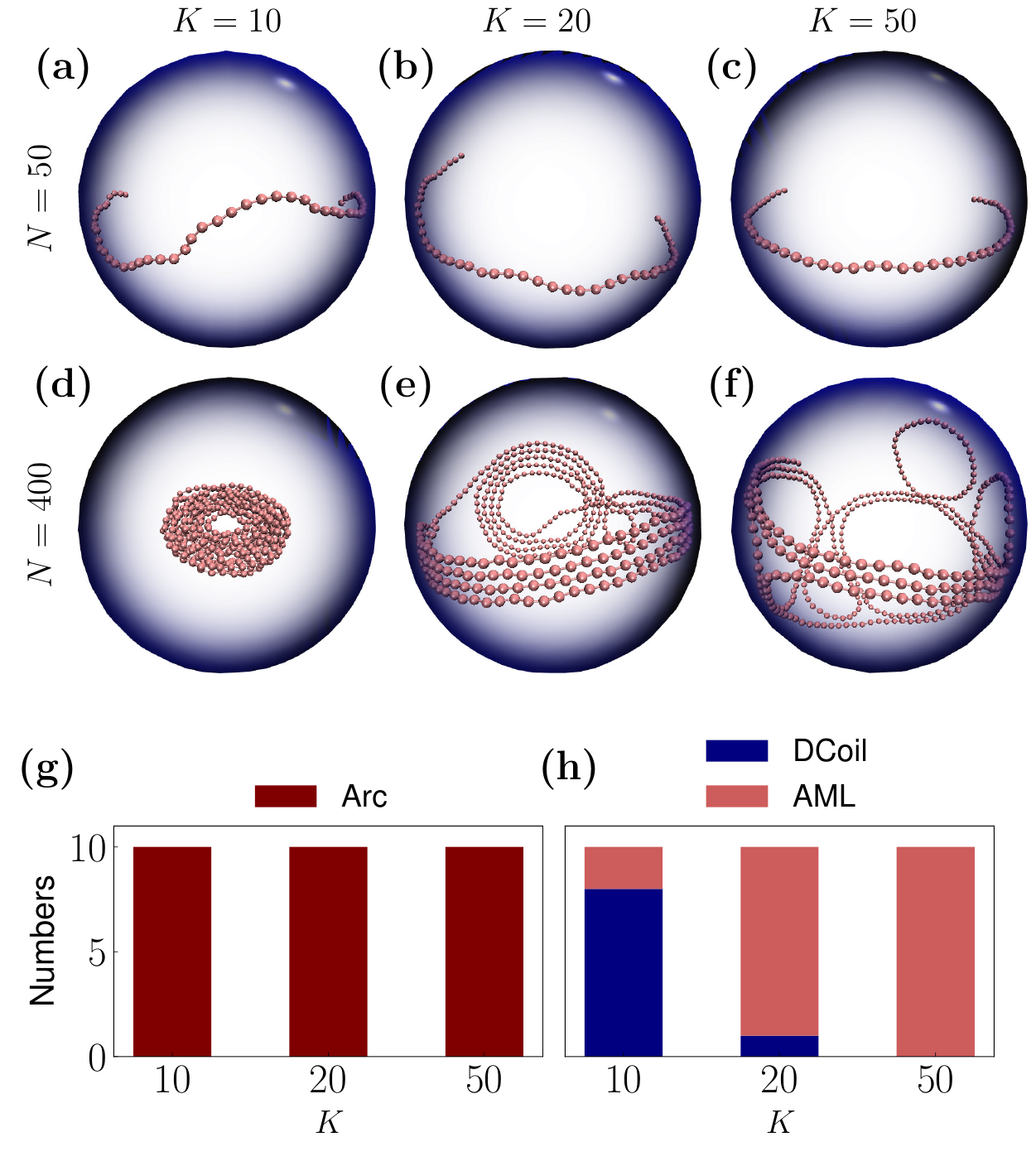}
\caption{Some example conformations for smaller ($N = 50$) and larger ($N = 400$) semiflexible polymers in presence of a surface; for different $K$ values, \textit{i.e.}, $K = 10, 20$ and $50$ respectively. For smaller $N(=50)$ but different $K$ values, the polymer assumes arc-like conformation as shown in \textbf{(a)}, \textbf{(b)} and \textbf{(c)}. In case of larger $N(=400)$, when $K$ is small ($K = 10$), the conformation is desorbed coil (DCoil) (see \textbf{(d)}) while for higher bending rigidities ($K = 20, 50$), the polymer assumes an adsorbed multiloop (AML) conformation as represented in \textbf{(e)} and \textbf{(f)} respectively. The number distribution of conformations for \textbf{(g)} $N = 50$ and \textbf{(h)} $N = 400$ for different $K$ values.}
\label{fig:conf-example}
\end{figure}

A fully flexible polymer in poor solvent conditions adopts a collapsed conformation under periodic boundary conditions. However, we have shown earlier~\cite{tripathi2019confined} that near attractive surfaces and in the presence of explicit crowders, the polymer adopts an adsorbed-layered (AL) configuration and transitions to an adsorbed-collapsed (AC) conformation, which depends on the strength of the monomer-surface attraction and the crowder density. This scenario is expected to change significantly with the introduction of bending rigidity $K$ in the polymer. For higher $K$ values, the polymer is not able to collapse even in poor solvent conditions due to the additional angle potential energy. Additionally, the introduction of a spherical surface for confinement in the semiflexible polymer system, makes the polymer assume a variety of conformations. There is an evident competition between the poor solvent condition and bending rigidity of the polymer in which the former has a tendency to collapse the polymer and the latter has a tendency to extend it. Further, the amount of extension of the polymer is limited by the size of the confinement and therefore, the conformations of the polymer also depend upon the ratio of the length of the polymer $N$ to the size of the confinement $R_c$. For a small semiflexible polymer, in the presence of a relatively larger spherical confinement ($N/R_c \approx 3.33$), the polymer forms an extended and adsorbed arc-like conformations (see Fig.~\ref{fig:conf-example} (a), (b) and (c)) for all values of $K$. Earlier studies have also demonstrated similar spiral conformations under strong 2D confinement~\cite{liu2008shapes}.

As the size of the semiflexible polymer increases, for a fixed confinement size, the polymer assumes more complex conformations. For the larger $N/R_c$ ratio $\approx 26.66$, instead of simple arc-like conformations, polymer takes a doughnut-like desorbed-coiled (DCoil) conformation (Fig.~\ref{fig:conf-example}(d)) for low $K$ values (\textit{i.e.} $ K = 10$). However, at higher $K$ values ($K > 10$), the polymer chain assumes an adsorbed-multiloop (AML) conformation as represented in Fig.~\ref{fig:conf-example}(e) and (f). Figure~\ref{fig:conf-example}(g) and (h) show the histograms of the conformations as a function of $K$ for two different values of polymer lengths $N = 50$ and $N = 400$. The data used to produce the histogram is sampled from $10$ independent simulations. The AML conformations in our study are closely related to helical conformations seen in earlier work by Yang \textit{et al.}~\cite{yang2011local}. They showed that a coil-helix transition occurs in semiflexible polymer confined within a non-interacting sphere. In our study, the monomer-wall interaction is attractive which leads to the adsorption of monomers onto the wall leading to AML conformations. The addition of repulsive crowders to such conformations is studied next.

\subsection{Effects of Crowder Density}\label{sec:B}
\begin{figure}[h]
\center
\includegraphics[width=\linewidth]{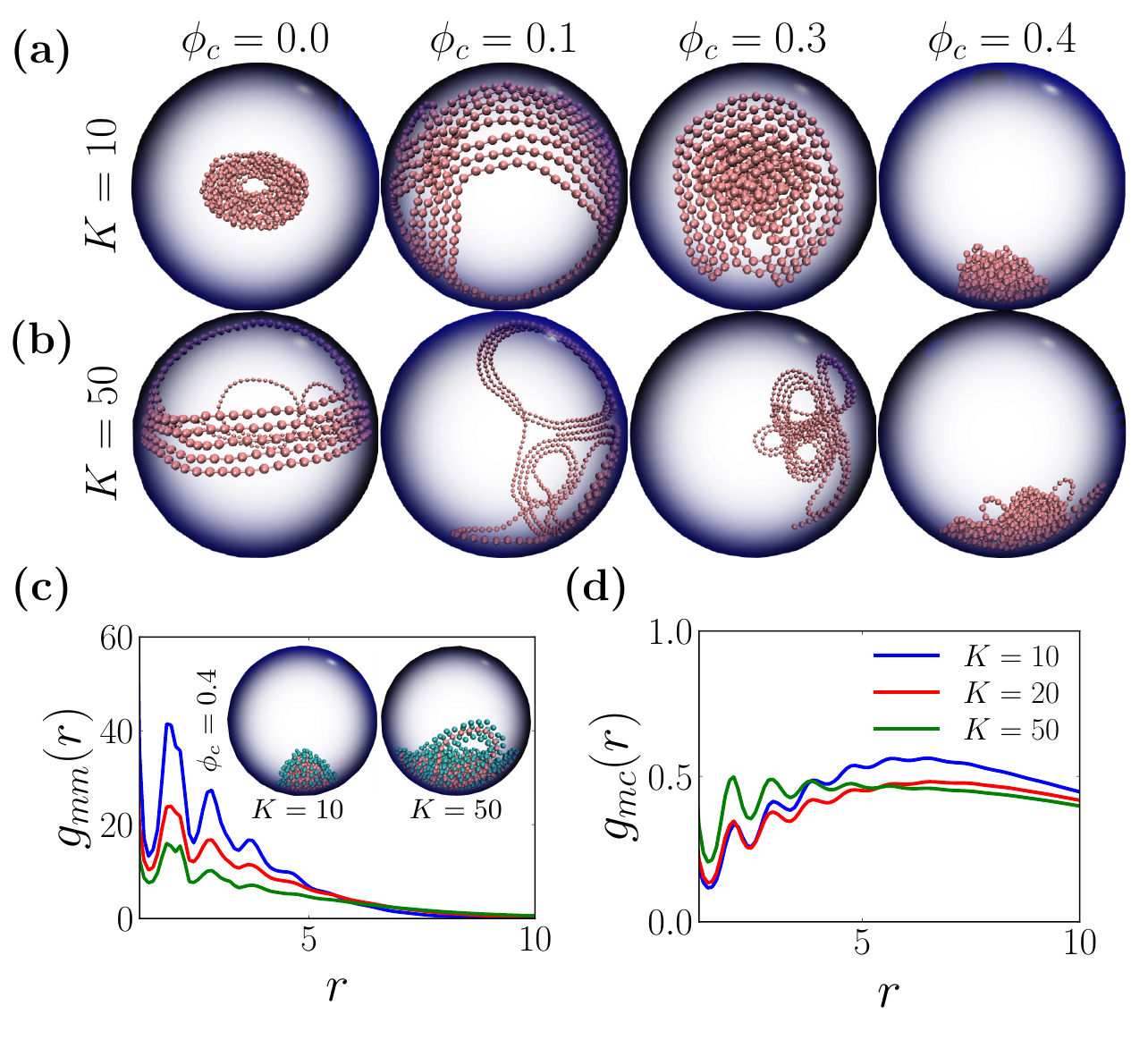}
\caption{Snapshots of semiflexible polymer ($N = 400$) for different crowder densities ($\phi_c = 0.1, 0.2, 0.3, 0.4$) at bending rigidity \textbf{(a)} $K = 10$ and \textbf{(b)} $K = 50$. The pair radial distribution function for \textbf{(c)} monomer-monomer pair $g_{mm}(r)$, and \textbf{(d)} for monomer-crowder pair $g_{mc}(r)$ at the highest crowder density ($\phi_c = 0.4$). Both plots represent the same color coding for $K$ values. Inset of \textbf{(c)} shows the snapshot of the polymer and the crowders within $1.5\sigma$ of a monomer for $K = 10, 50$ respectively at $\phi_c = 0.4$. The green and pink color beads represent the crowders and the monomers respectively.}
\label{fig:conf-varyPhi-gofr}
\end{figure}
In this section, we explore the effects of crowders on the conformations of a long neutral semiflexible polymer($N=400$). For smaller $K$ ($=10$) values, in the absence of crowders ($\phi_c = 0$), the polymer forms a DCoil conformation as discussed in sec.~\ref{sec:A}. With the addition of repulsive crowders in the system, the polymer coats the surface of confining surface and undergoes conformational changes with increasing crowder density:  an adsorbed-layered conformation (AL) at moderate crowder densities, and eventually collapses to an adsorbed-collapsed (AC) conformation at the highest crowder density ($\phi_c = 0.4$) as shown in Fig.~\ref{fig:conf-varyPhi-gofr} (a). With increase in $K$ value to $50$, the polymer undergoes adsorbed-multiloop (AML) conformation at low crowder densities ($\phi_c = 0-0.3$) to eventual AC conformation. For both flexibility values of the polymer, the higher crowder(repulsive) densities induce an adsorbed collapsed conformation due to depletion interactions(see Fig.~\ref{fig:conf-varyPhi-gofr} (b) from left to right). To understand the internal structure of the collapsed conformations as a function of the semiflexibility of the polymer, the pair radial distribution function between monomers, $g_{mm}(r)$, is computed for the highest crowder density. For low bending rigidity ($K = 10$), the polymer adopts a very compact AC conformation in which the average number of nearby monomers is high (refer to the inset of Fig.~\ref{fig:conf-varyPhi-gofr}\textbf{(c)}, $K = 10$) and results in well-ordered peaks in $g_{mm}(r)$. This gives rise to a pronounced peak in $g_{mm}(r)$ as plotted in Fig.~\ref{fig:conf-varyPhi-gofr} (c). As $K$ increases, the competition between the depletion interactions due to crowders and the increased stiffness potential energy results in a less compact AC conformation, as can be seen by the loss of well-defined peaks in $g_{mm}(r)$ (refer to the inset of Fig.~\ref{fig:conf-varyPhi-gofr}(c), $K = 50$). Figure~\ref{fig:conf-varyPhi-gofr}(d) shows the pair radial distribution $g_{mc}(r)$ for monomer-crowder pair. The emergence of more open structures with increasing polymer stiffness is reflected as increased monomer-crowder interactions with trends opposite to that of Fig.~\ref{fig:conf-varyPhi-gofr}. A more open conformation of polymer leads to enhanced monomer-crowder interactions which can be seen in Fig.~S1(a). It is seen that the number of crowders within $1.5\sigma$ distance of the polymer is higher for $K=50$ than that for $K=10$. We also characterize the AC conformation of the polymer at $\phi_c=0.4$ via the radial density function $\rho(r')$, which represents the monomer density as a function of distance from the center of the confinement along the radial direction. Figure~\ref{fig:rho-varyK} shows the $\rho(r')$ for different bending rigidities at the highest crowder density $\phi_c = 0.4$. The results show that the stiffer polymers have higher adsorption on the confining wall as compared to the more flexible polymers, consistent with previous studies ~\cite{sintes2001adsorption, kramarenko1996molecular,chauhan2021adsorption}.

We conclude that though there are differences in the internal structures of adsorbed collapsed conformations of polymers with different chain flexibility, the highest crowder density, $\phi_c=0.4$, considered in this study does induce a collapsed conformation, dramatically different from $\phi_c=0.3$. We note that in our earlier study of long completely flexible polymer~\cite{tripathi2019confined}, we observe a similar collapsed or crumpled polymer conformations and this underscores the role of crowder densities vis-a-vis polymer flexibility: high crowder densities effectively renormalize the flexibility of the polymer. 
\begin{figure}[h]
\center
\includegraphics[width=\linewidth]{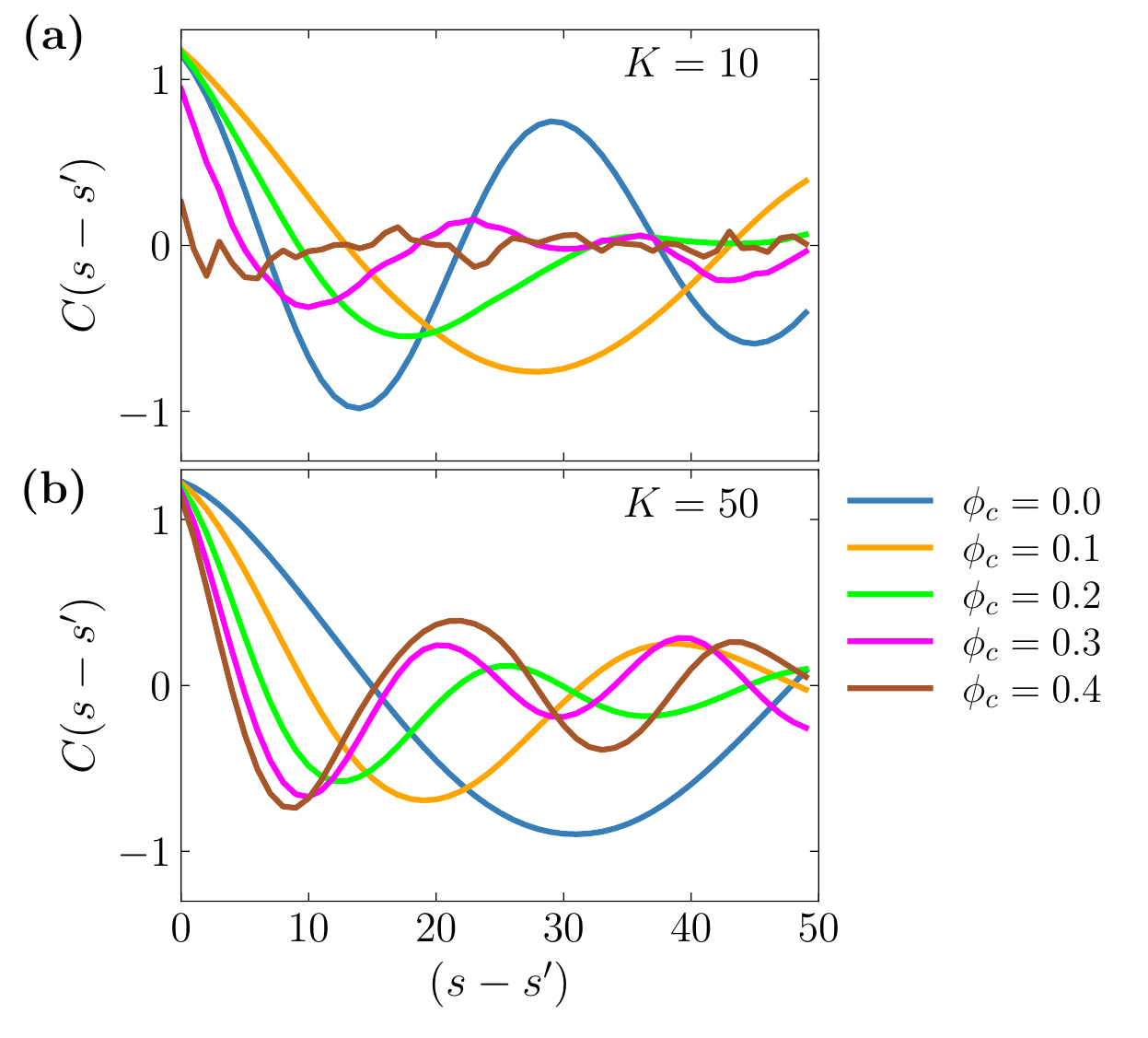}
\caption{Tangent-tangent correlation function $C(s - s{'})$ as a function of contour distance $s-s{'}$ for \textbf{(a)} $K = 10$ and \textbf{(b)} $K = 50$ for different crowder densities $\phi_c$ for a linear polymer of length $N = 400$ in poor solvent condition.}
\label{fig:bibjplots}
\end{figure}

To quantify the renormalization of the polymer flexibility as a function of crowder density, we calculate the tangent-tangent bond correlation function
\begin{equation}
C (s - s{'}) = \langle \mathbf{b}(s) . \mathbf{b}(s{'})  \rangle = \langle \cos \theta (s-s{'})  \rangle, 
\end{equation}

with $\mathbf{b}(s)$ being the bond vector at site $s$ and $\theta(s-s{'})$ being the angle between the tangents to the polymer at site $s$ and $s{'}$. In good solvent and unconfined condition, $C (s - s{'})$ is expected to decay exponentially at the length scale $l_p$ as the directions of parts of the polymer chain become uncorrelated as $s-s{'} >> l_p$, where, $l_p$ is the persistence length of the semiflexible polymer and is related to the bending rigidity with $l_p = K/k_B T$. However, in poor solvent conditions or confined scenarios, $C(s-s{'})$ does not show an exponential decay, but an oscillatory behavior. Similar oscillatory behavior in $C(s-s{'})$ has been observed in actin filaments confined in narrow channels~\cite{koster2005brownian, koster2007fluctuations} showing a deviation from a bare persistence length~\cite{liu2008shapes, odijk1983statistics}. The length scale of oscillations is related to the size of the loop formed by the polymer~\cite{yang2011local}. In order to calculate $C(s - s{'})$, the dot product of two bond vectors separated by distance $s-s{'}$ is computed and averaged over the chain length with $s-s{'}$ varying from $0$ to $50$. Figure~\ref{fig:bibjplots}(a) shows $C(s - s{'})$ for $K = 10$ at different crowder densities. 

In the case of DCoil conformation which corresponds to the $K=10$ and $\phi_c = 0.0$, the tangents get decorrelated at a smaller length scale than that in AML conformations which correspond to $K=50$ and $\phi_c = 0.1-0.3$, as the size of loop formed in DCoil is smaller than that of AML. As the crowder density increases, at $K = 10$, the loop structure, as well as the oscillations, disappear in $C(s - s{'})$ curve as shown in Fig.~\ref{fig:bibjplots}(a) ($\phi_c = 0.4$). At $K = 50$, the loop structure exists regardless of crowder density, however, the loop size becomes smaller with crowder density. This can be seen in the Fig.~\ref{fig:bibjplots}(b) as the correlation curves become steeper with the crowder density, strongly indicating that the effective persistence length decreases with increasing crowder density.

%\begin{equation}
%C(s - s{'}) \sim  \exp(s-s{'})/l_p , 
%\end{equation}
\begin{figure}[h]
\center
\includegraphics[width = 0.8\linewidth]{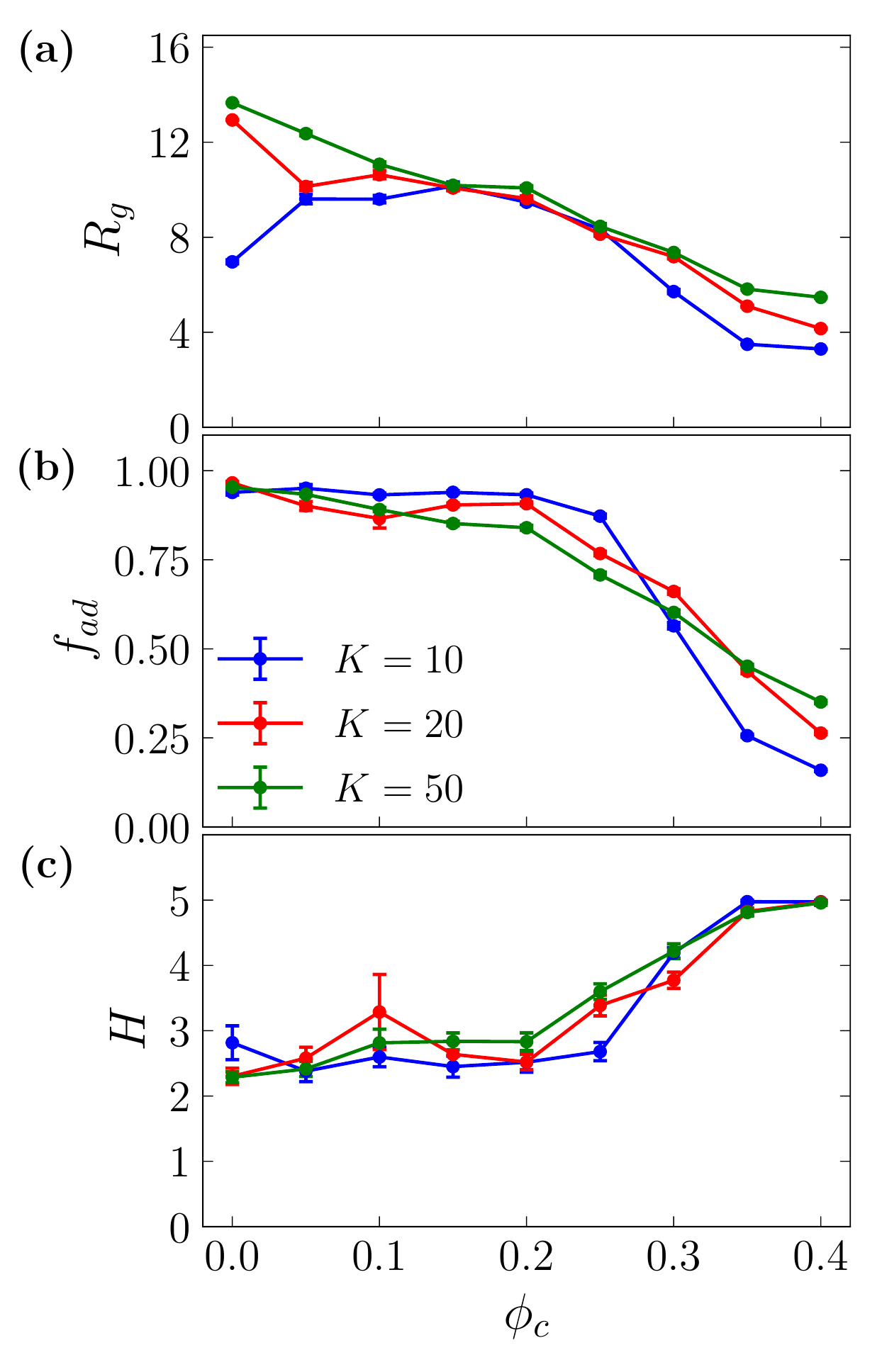}
\caption{\textbf{(a)} The radius of gyration ($R_g$), \textbf{(b)} the fraction of adsorbed monomers ($f_{ad}$) and \textbf{(c)} height of the adsorbed polymer $H$ is plotted against $\phi_c$ for different bending rigidities $K$ for linear polymer of size $N=400$. The same color coding applies for $K$ values in all the subplots.}
\label{fig:rg-fad-h}
\end{figure}

We now analyze additional parameters related to the polymer's adsorption on the wall, such as the radius of gyration $R_g$, the fraction of monomers adsorbed onto the surface $f_{ad}$, and the height of the adsorbed polymer $H$. The radius of gyration $R_g$ which provides a general idea of the overall size of the polymer, and, is given by the following expression,
\begin{equation}
R_g ^2 = \frac{1}{2N^2}~\sum_{i=1}^N \sum_{j=1}^N |({\vec r}_i - {\vec r}_j)|^2,
\label{eq:rgDef}
\end{equation}
where $r_i$ and $r_j$ represent the position vectors of two monomers $i$ and $j$ respectively. For an adsorbed conformation (AC), we calculate the fraction of monomers adsorbed onto the surface $f_{ad}$, which is defined as the ratio of the number of monomers adsorbed onto the surface to the total number of monomers of a polymer chain. A polymer is considered in adsorbed conformation only if at least $5$ of its monomers are adsorbed onto the confining wall. A monomer is considered adsorbed if the distance of the center of mass of the monomer lies within $1.5\sigma$ of the wall of the confinement. We also compute a height function for the adsorbed conformations. The height function measures the height of the adsorbed conformation relative to the confining sphere surface. To calculate the height of the adsorbed polymer conformation, we calculate the distance of each monomer from the center of the confining sphere and subtract this distance from the radius of the confinement $R_c$. The largest number thus obtained is a measure of the height of the AC conformation and is termed $H$. The $f_{ad}$ and $H$ are averaged over the last $10^6$ steps of the MD trajectory for all the adsorbed conformations out of total $10$ independent simulations. 

There is an overall reduction in the dimension of the adsorbed polymer as the crowder density $\phi_c$ is increased (see Fig.~\ref{fig:rg-fad-h}(a)) for different values of $K$.  For low values of $K$, as the crowders are introduced in the system, the radius of gyration $R_g$ first increases and then decreases ($K = 10$). The reason for this non-monotonic behavior can be explained as follows: for the lower value of $K$ in the absence of crowders, the polymer assumes DCoil conformation which is relatively compact and has a lower $R_g$, but as $\phi_c$ is increased, the polymer adsorbs onto the wall and manifests an AL conformation which has a higher $R_g$, further increment in $\phi_c$ makes the polymer crumple and take an AC conformation, hence $R_g$ showing the non-monotonic behavior. However, for higher values of $K$, the polymer is in AL conformation, and increasing the crowder density leads to $R_g$ values decreasing monotonically. With the increase in crowder (repulsive) density, the free energy of the system is lower when the polymer minimizes its exposure to the crowders and leads to more compact conformations. This is also reflected in the  $f_{ad}$ values, which capture the number of monomers in contact with the confining surface, which shows a sharp decrease for $\phi_c >=0.3$(see Fig.~\ref{fig:rg-fad-h} (b)). At these higher crowder densities, the polymer starts to protrude in the radial direction while compacting in tangential directions giving rise to the concomitant rise of the height of the adsorbed conformation $\phi_c$ as seen in Fig.~\ref{fig:rg-fad-h} (c).

\subsection{Effects of Bending Rigidity}\label{sec:C}
We explore the effects of bending rigidity on the polymer conformations in this section. It is expected that as the bending rigidity is increased, the polymer will assume extended states which will lead to better adsorption and larger size.
We plot the radius of gyration $R_g$ as a function of bending rigidity $K$ for different crowder densities in Fig.~S2(a). For all $\phi_c$ values, $R_g$ increases initially as $K$ increases and then saturates to a value. This saturation of $R_g$ can be explained by the fact that polymer extends maximally under confinement to moderately higher values of $K$ and further rising $K$ does not lead to the increment of the $R_g$. This saturation value is dictated by the size of the confining sphere $R_c$ and the crowder density $\phi_c$. To understand the effects of bending rigidity on the adsorption of the polymer, we plot the fraction of adsorbed monomers to the surface $f_{ad}$ with $K$ in Fig.~S2(b). For lower values of $\phi_c$ ($= 0, 0.1, 0.2$), almost all the monomers adsorb on the wall ($f_{ad} \approx 1$) and increment in $K$ does not make any significant difference to $f_{ad}$. However, at higher values of $\phi_c$, as the polymer is in AC phase and increasing $K$ leads to a slight increase in $f_{ad}$. We also plot the height $H$ of the adsorbed polymer in Fig.~S2(c). Similar to $f_{ad}$, the height shows no difference with $K$ for low crowder densities. At $\phi_c = 0.4$, the average height remains constant regardless of $K$. 
 
\begin{figure}[H]
\center
\includegraphics[width=0.8\linewidth]{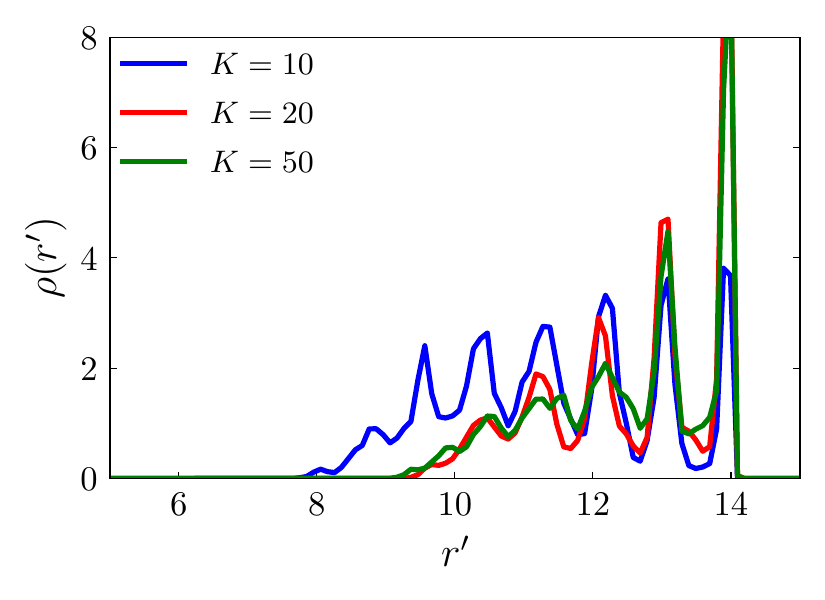}
\caption{The radial density function $\rho(r')$ of monomers at the highest crowder density ($\phi_c = 0.4$) for bending rigidities $K=10, 20, 50$.}
\label{fig:rho-varyK}
\end{figure}

The radial density function $\rho(r')$ represents the monomer density as a function of distance from the center of the confinement along the radial direction. Figure~\ref{fig:rho-varyK} shows the $\rho(r')$ for different bending rigidities at the highest crowder density $\phi_c = 0.4$. For lower $K$ values, the polymer is in AC phase. Therefore, the radial density function's peaks are spread around the wall (\textit{i.e.} $r^{\prime} \approx 15$). As the value of $K$ increases, the adsorbed polymer starts coating the wall which can be seen in the pronounced peaks of $\rho(r^{\prime})$ near the confining wall.

\subsection{Closed topology restricts the variety of the conformations}\label{sec:D}
In this section, we study the effects of closed topology on the conformations of a polymer. To understand the differences between the conformations of a polymer with an open and closed topology, we simulated a ring polymer of same length ($N = 400$) under the same conditions as the linear polymer and compared the results.

\onecolumngrid

\begin{figure}[ht]
\includegraphics[width = 0.9\linewidth]{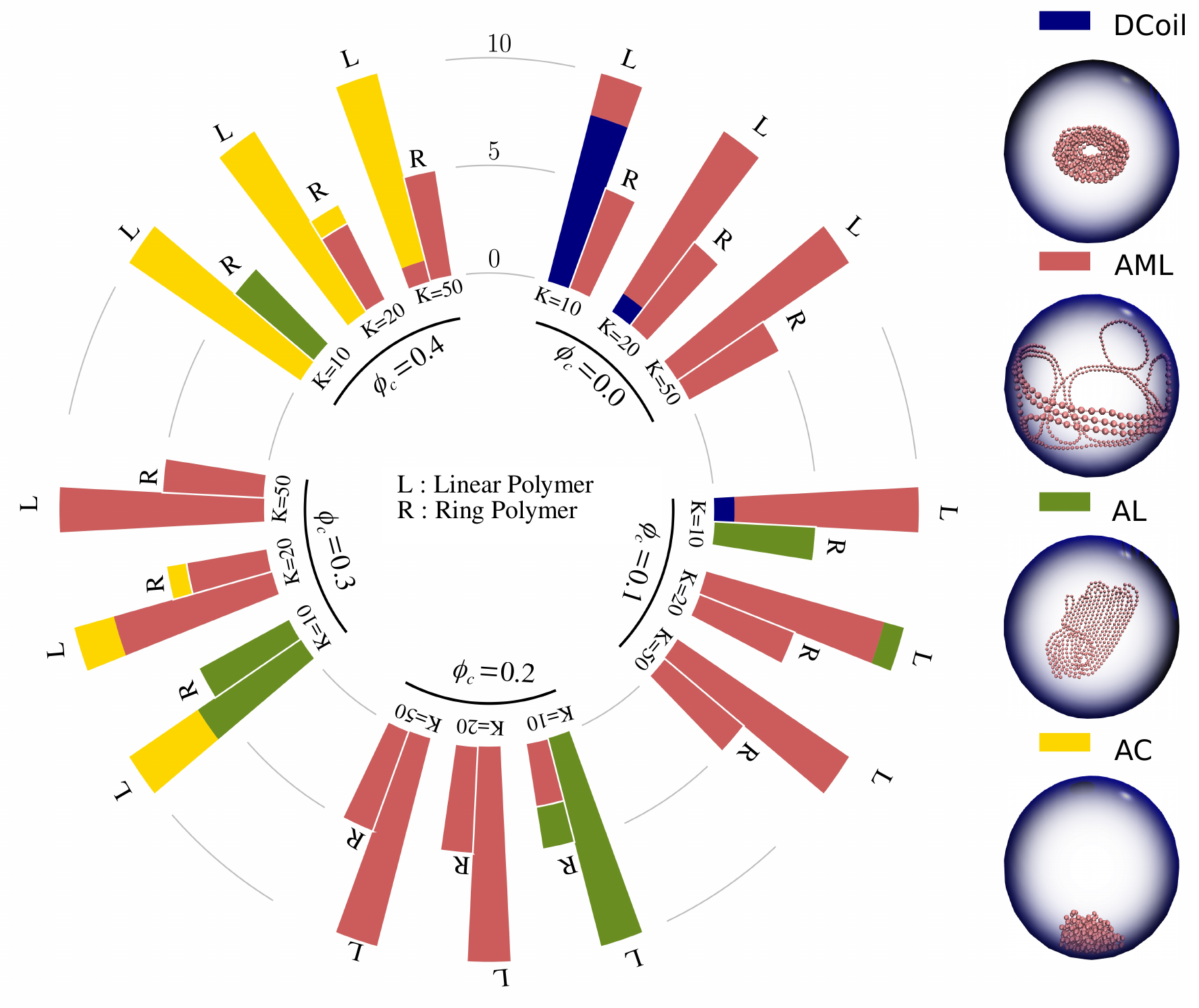}
\caption{Comparative bar plot of the emergent conformations  for linear (L) and ring (R) polymer of size $N=400$ at different crowder densities ($\phi_c$) and bending rigidity ($K$). Legends show snapshots of example conformations assumed by the category. DCoil, AML, AL and AC refer to desorbed-coiled, adsorbed-multiloop, adsorbed-layered and adsorbed-collapsed conformations respectively.}
\label{fig:circ_bar}
\end{figure}

\twocolumngrid

Figure~\ref{fig:circ_bar} shows the conformation occurance of a polymer. In general, for $N=400$, as we have discussed in earlier sections, there are four possible types of conformations: `DCoil', `AML', `AL', and `AC'. Depending upon the crowder density, bending rigidity, and topology, the polymer assumes one of these conformations. Our simulations strongly suggest that for ring polymers with closed topology, the probability of finding a ring polymer in AC state is extremely low. As discussed in sec.~\ref{sec:B}, at high density ($\phi_c = 0.4$), the linear polymer is in AC phase, however, at the same density, the ring polymer is found in either AL or AML conformation. In other words, an extra constraint on the polymer structure (\textit{i.e.}, closed topology) precludes the AC phase. Further, the linear polymer assumes, a DCoil phase at $\phi_c = 0.0$ and lower bending rigidity ($K=10$), which is also absent in ring polymers. The DCoil phase comes into existence, possibly due to the entanglement of the ends of the polymer, and, ring polymer having no ends, can not exhibit this phase. At moderate crowder densities, both ring polymer and linear polymer show similar conformations (see Fig.~\ref{fig:circ_bar}). 

\begin{figure}[h]
\includegraphics[width = 0.8\linewidth]{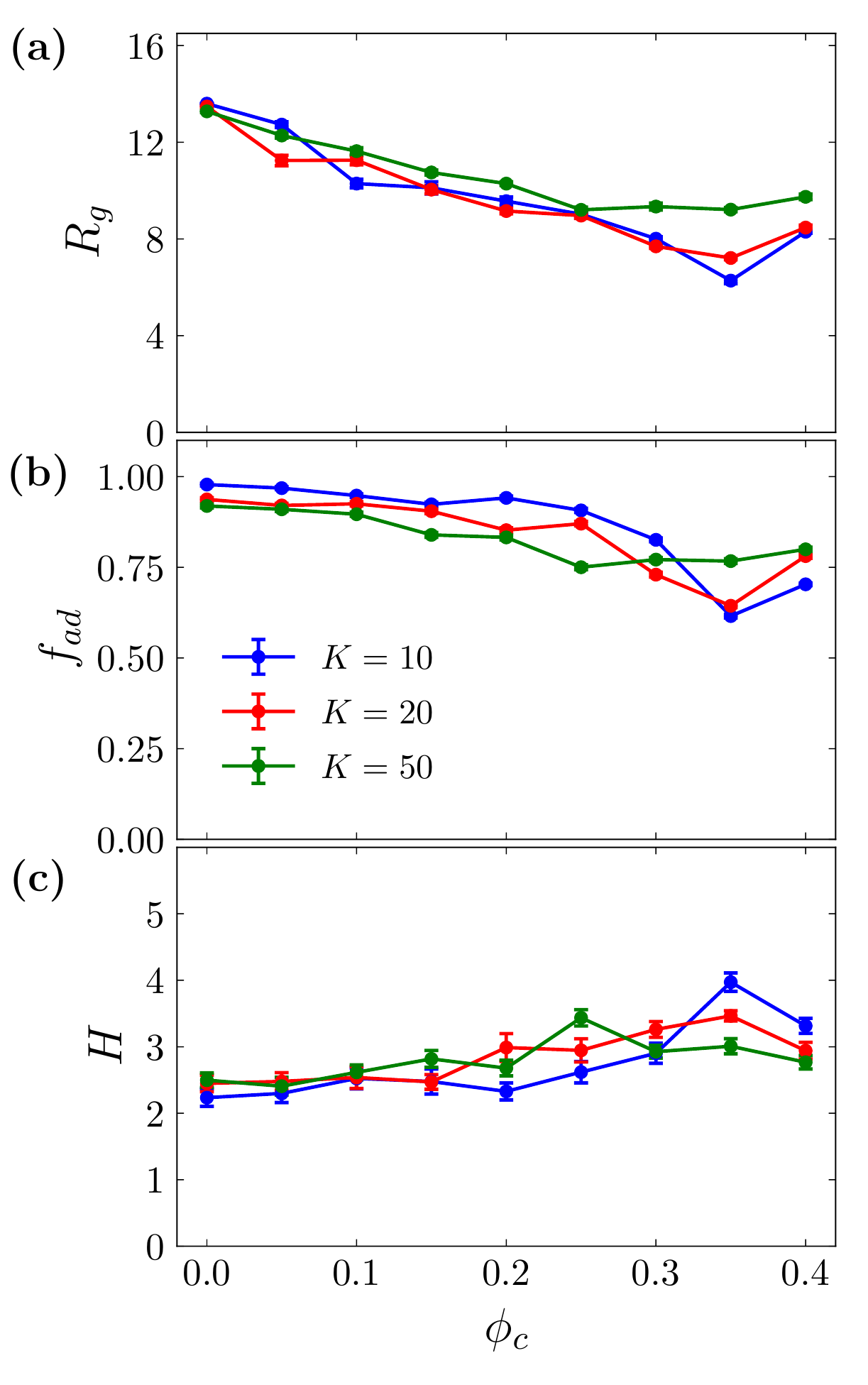}
\caption{\textbf{(a)} The radius of gyration $R_g$, \textbf{(b)} the fraction of adsorbed monomers $f_{ad}$ and \textbf{(c)} height $H$ of the adsorbed polymer is plotted against $\phi_c$ for different bending rigidities $K$ for ring polymer of size $N =400$. The same color coding applies for $K$ values in all the subplots.}
\label{fig:rg-fad-h-LinearvsRing}
\end{figure}

We also calculate parameters $R_g$, $f_{ad}$, and $H$ for the ring polymer and compared these with their counterparts in the linear polymer. $R_g$ of the ring polymer shows a decrease with the crowder density. At the highest crowder density, $\phi_c = 0.4$, $R_g$ is not as low as for linear polymer because of its inability to achieve the most compact AC conformation (see Fig.~\ref{fig:rg-fad-h}(a) and~\ref{fig:rg-fad-h-LinearvsRing} (a)). Due to the closed topology, the ring polymer does not collapse and it adsorbs better on the confinement surface than the linear polymer. From Fig.~\ref{fig:rg-fad-h}(b) and~\ref{fig:rg-fad-h-LinearvsRing} (b), it can be seen that $f_{ad}$ is in general higher from ring polymer indicating that the monomers of a relatively open ring polymer easily adhere to the wall than of a collapsed linear polymer. The fact that most of the monomers of a ring polymer adhere to the wall of confinement also reflects in the height of the adsorbed polymer (see Fig.~\ref{fig:rg-fad-h} (c) and~\ref{fig:rg-fad-h-LinearvsRing}(c)). The difference between the polymer conformations in linear and ring cases can also be understood by comparing the number of crowders in contact with the polymers. The Fig.~S1 shows the number of crowders within $1.5\sigma$ distance of the polymers with linear (L) and ring (R) topology. The higher number of crowders in contact with the ring polymer suggests that for the same backbone flexibility, the effect of crowder density on the effective persistence length of linear polymer is higher compared to that of a ring polymer. 

%%%%%%%%%%%%%%%%%%%%%%%%%%%%%%%%%%%%%%%%%%%%%%%%%%%%%%%%%%%%%%
%%%%%%%%%%%%%%%%%%% DISCUSSION %%%%%%%%%%%%%%%%%%%%%%%%%%%%%%%%%%%
%%%%%%%%%%%%%%%%%%%%%%%%%%%%%%%%%%%%%%%%%%%%%%%%%%%%%%%%%%%%%%
\section{Discussion and Conclusion \label{sec:conclusions}}
In this paper, we use large-scale molecular dynamics simulations to study the conformational landscape of a polymer under the competing effects of bending rigidity and the crowder density under confinement. The solvent condition of the system is kept poor in all the simulations while the bending rigidity and the crowder density are varied. We also compared the conformations of a linear and ring polymer. The formation of loops in semiflexible polymers is of particular interest in cases such as the binding of transcription factors on DNA via the formation of loops~\cite{han2009concentration}. An intriguing aspect of the semiflexibility of DNA molecules is that the predicted persistence length of the DNA ($\approx 50$ nm) from theoretical models such as worm-like chain models~\cite{kratky1949rontgenuntersuchung,baschnagel2016semiflexible} cannot explain the compact conformations of DNA around histone proteins~\cite{prinsen2010nucleosome}. This strongly suggests that the crowded environments under which DNA exists may play a vital role in effectively modifying the semiflexibility of such biopolymers.  Two-dimensional semiflexible polymers in a  hard-disk fluid have been shown, using simulations, to display strong renormalized persistence lengths~\cite{schobl2014persistence}. In particular, that study predicts that the effective persistence lengths may be reduced in a crowded environment due to the crumpling of the semiflexible polymers due to crowding. There are experimental studies showing that macromolecular crowding affects the rigidity of DNA~\cite{zhang2009macromolecular,nir2011hu}.

\begin{figure}[ht]
\centering
\includegraphics[width = 0.9\linewidth]{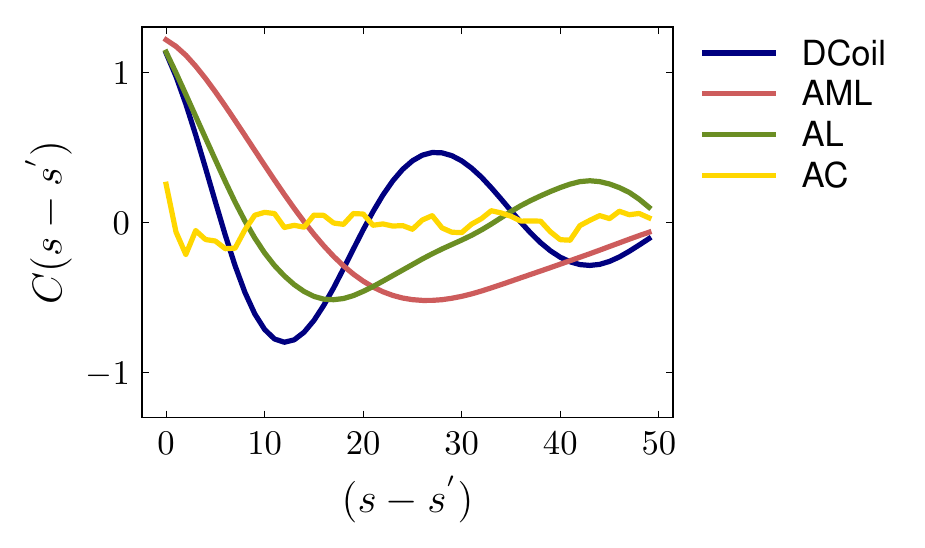}
\caption{The tangent-tangent correlation function $C(s-s{'})$ for desorbed-coiled (DCoil), adsorbed-multiloop (AML), adsorbed-layer (AL) and adsorbed-collapsed (AC) conformations.}
\label{fig:bibj-diffconf}
\end{figure}

Theoretical works on the adsorption of semiflexible polymers near a flat attractive surface~\cite{kuznetsov1998semiflexible} suggest a desorbed to a pancake-like adsorbed conformation for higher values of polymer stiffness. However, if smaller polymer lengths are considered, it may not be possible to observe the pancake-like phase, as that would require longer polymer lengths. Earlier works on the simulation of short polymers ($N \approx 50$) predict a simple conformational phase diagram for semiflexible polymers near surfaces including lines, trains, and loops as possible conformations of adsorbed semiflexible polymers. However, the length of the semiflexible polymer can have significantly richer conformational phases as seen in our study. Many earlier works suggested that semiflexible polymers absorb more easily onto surfaces than flexible polymers due to comparatively lower entropy loss when adsorbed on surfaces. In the presence of crowder molecules, confining conditions, and attractive surfaces, our results corroborate the same for linear semiflexible polymers as well. However, in the case of topologically constrained large semiflexible ring polymers, the surface adsorbing modes are different compared with linear semiflexible polymers. Linear semiflexible polymers are seen to undergo higher compaction, in the presence of crowders, as compared to ring semiflexible polymers. This suggests that the effective renormalization of the persistence lengths by the crowders affects linear and ring polymers in very different ways. In the absence of crowders, a small polymer takes an arc-like confirmation for all $K$ values, however, a larger polymer assumes DCoil and AML conformation for lower and higher $K$ values respectively. As the crowders are introduced into the system, the small polymer assumes hairpin-like or spiral-like conformations while a larger polymer coats the confinement wall, and a further increase in the crowder density leads to an AC conformation. At smaller bending rigidity, since the bending energy component is small, the polymer can easily bend and goes from DCoil $\rightarrow$ AL $\rightarrow$ AC conformation with the increase in crowder density. However, at higher values of the bending rigidity, the polymer resists collapsing and assumes AML $\rightarrow$ AC as the crowder density increases.The tangent-tangent correlation function $C(s-s{'})$ are very distinct in all the conformations (see Fig.~\ref{fig:bibj-diffconf}).

We also investigate the effects of linking the two ends of a polymer. It is found that the conformational landscape of a ring and a linear polymer are identical at moderate crowder density for large polymers. However, the difference emerges for the smaller polymer's conformations. In case of open topology, the polymer takes a diverse range of conformations while in case of closed topology, the diversity of conformations, a polymer can adopt, are restricted. This is further corroborated by the data that shows that ring polymers exhibits larger radius of gyration, higher adsorption and lower height at higher crowder densities than their linear counterparts (see Fig.~S3). In a biological setting, polymers are surrounded by particles of many different sizes. However, in this study, the size of all the crowder particles is kept same and the effects of polydispersity is neglected. 
Introducing a polydisperse set of crowders may lead to interesting effects on the conformation of the polymers. Along with that it is also notable that the adsorption of the polymer onto the wall may deform the wall in real systems. Introducing a flexible wall rather than a rigid one will lead us to closer to a more realistic systems.  These are all a part of our future studies.

\end{document}

\title{\Huge \underline{Supplementary Information}\\ 
{\huge {Conformational landscape of long semiflexible linear and ring polymers near attractive surfaces}} \\
}
\author[]{Kamal Tripathi} 
\affil[]{The Institute of Mathematical Sciences, C.I.T. Campus, Taramani, Chennai 600 013, India}
\author[]{Satyavani Vemparala} 
\affil[]{The Institute of Mathematical Sciences, C.I.T. Campus, Taramani, Chennai 600 013, India}
\affil[]{Homi Bhabha National Institute, Training School Complex, Anushaktinagar, Mumbai 400 094, India}
\date{} \maketitle \newpage

\section{Effects of topology}
\begin{figure}[ht]
\includegraphics[width =\linewidth]{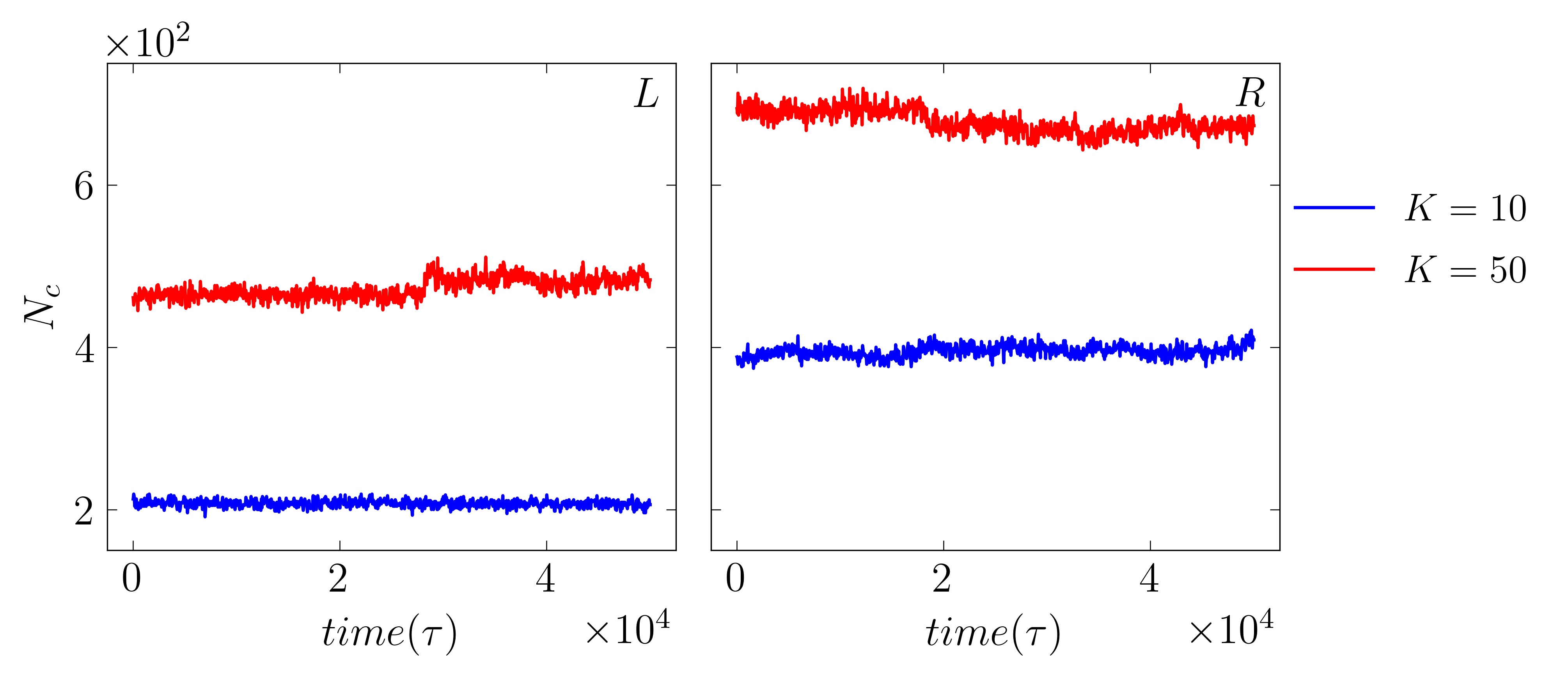}
\caption{The number of crowders $N_c$ within the distance $1.5 \sigma$ of polymer is plotted with time at different bending rigidities $K = 10, 50$ in the highest density cases ($\phi_c = 0.4$) for \textbf{(a)} Linear Polymer $N = 400$, and \textbf{(b)} Ring Polymer $N = 400$.}
\label{fig:Nc-LinearvsRing}
\end{figure}

The Fig.~\ref{fig:Nc-LinearvsRing} shows the number of crowders $N_c$ within $1.5\sigma$ distance of the polymer in highest density case ($\phi_c=0.4$). The number $N_c$ is lower for smaller bending rigidity $K=10$ and higher for $K=50$ indicating the fact that at higher bending rigidity, the polymer does not collapse and has more contact with the crowders when compared to the lower $K$ value. It also shows that the ring polymers, due to their topology, do not collapse when compared to their linear counterparts under the same conditions.

%%%%%%%%%%%%%%%%%%%%%%%%%%%%%%%%%%%%%%%%%%%%%%%%%%%%%%%%%%%%%%%%%%%%%55
\section{Effects of Bending Rigidity}
\begin{figure}[ht]
\center
\includegraphics[width=0.8\linewidth]{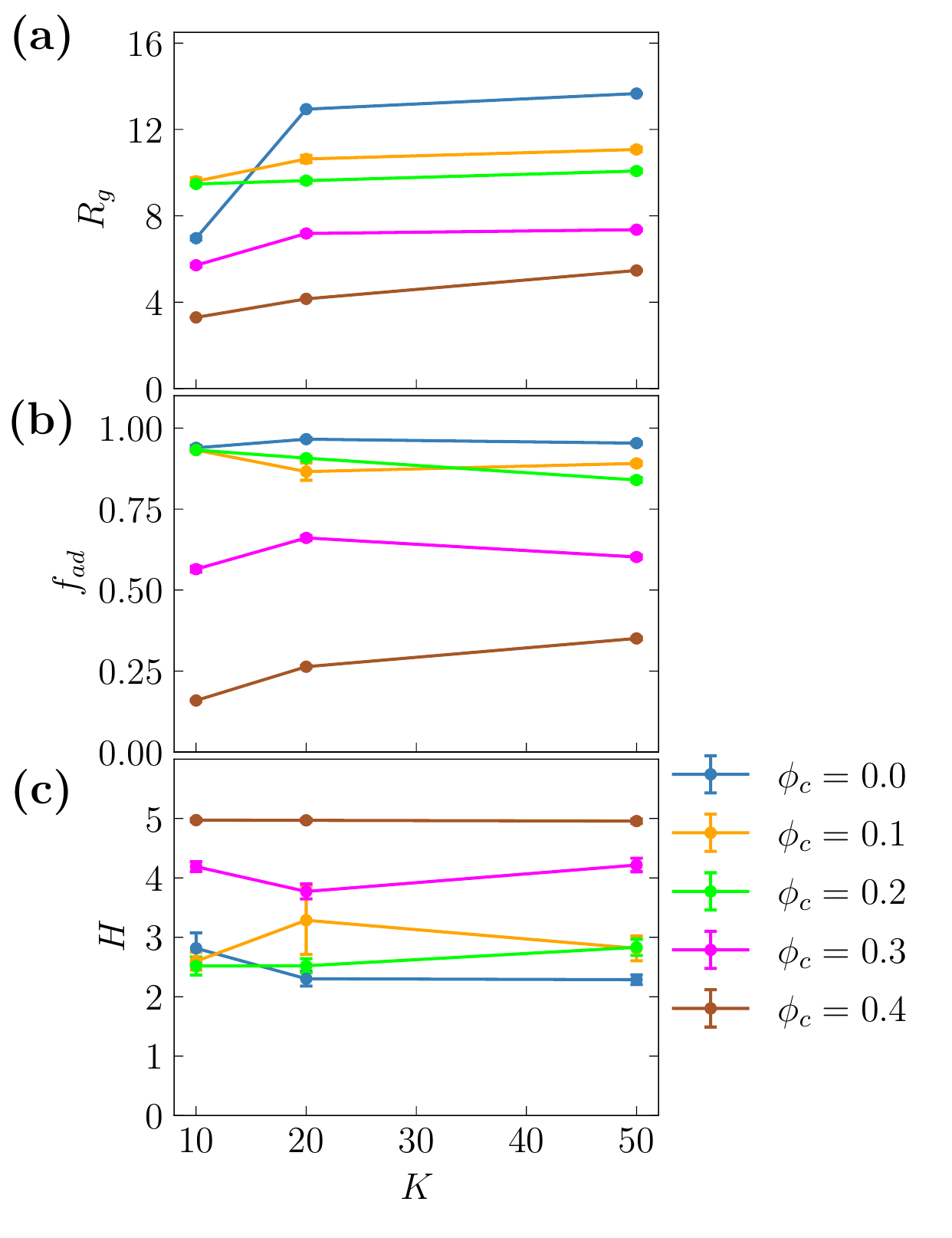}
\caption{Variation of $R_g$, $f_{ad}$ and $H$ with the bending rigidities $K$ for different values of $\phi_c$ for linear polymer of length $N = 400$.}
\label{fig:rg-fad-H-vsK}
\end{figure}

\begin{figure}[ht]
\center
\includegraphics[width=0.8\linewidth]{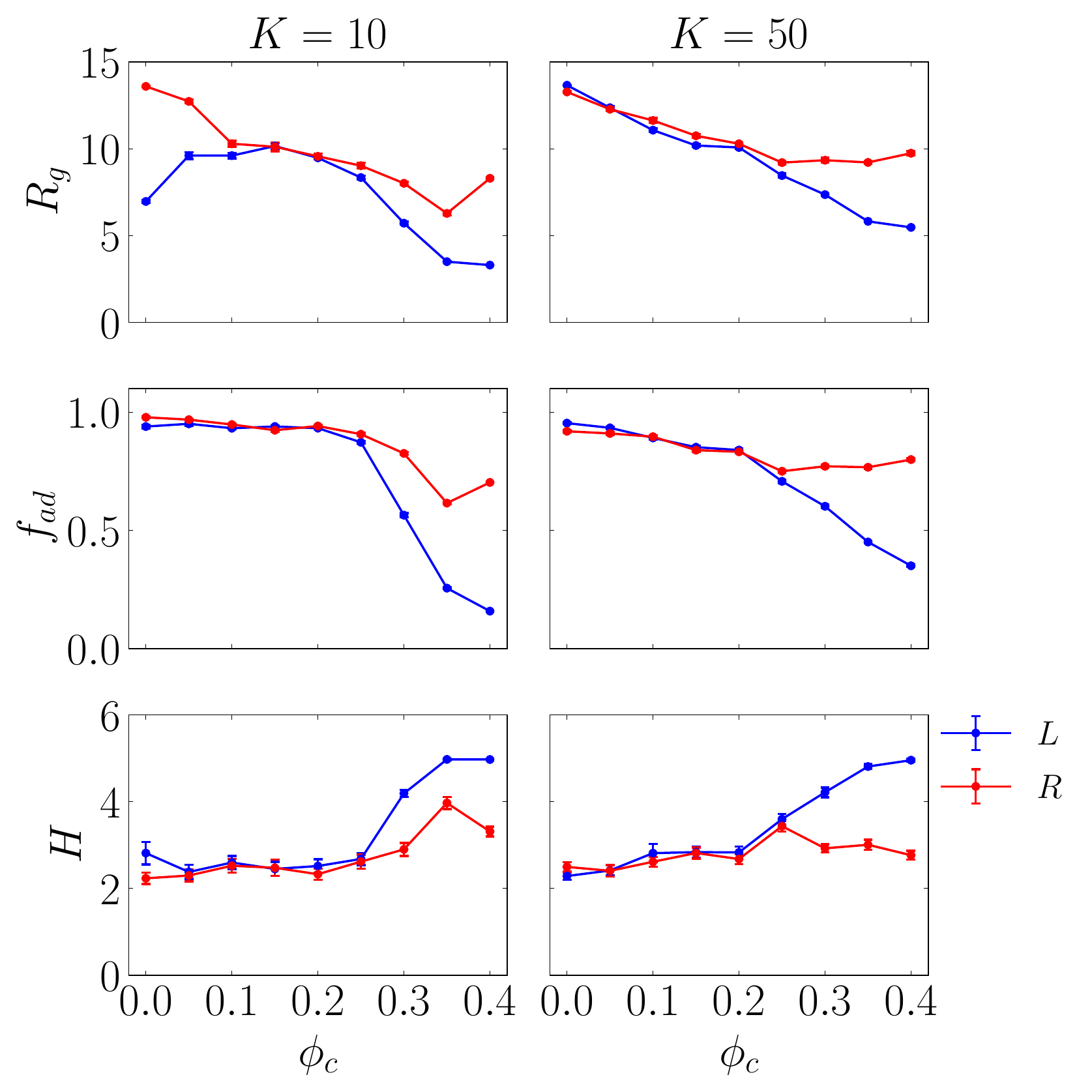}
\caption{Variation of $R_g$, $f_{ad}$ and $H$ with the crowder density $\phi_c$ at bending rigidities $K = 10, 50$ for linear (L) and ring (R) polymer of length $N = 400$.}
\label{fig:rg-fad-H-vsphi_LR}
\end{figure}

\clearpage